\documentclass[slac_one]{revtex4}
\usepackage{graphicx}
\usepackage{fancyhdr}
\usepackage{psfrag}

\usepackage{fancybox}
\usepackage{rotating}
\usepackage{amsmath}
\usepackage{exscale}
\usepackage{pst-node}
\usepackage{graphics}
\usepackage{euscript,array}
\usepackage{scalefnt}

\newcommand{\cE}{{\cal E}}
\newcommand{\cC}{{\cal C}}
\newcommand{\cR}{{\cal R}}
\newcommand{\barcC}{{\cal \bar C}}
\def\cO#1{{\cal O}\left( {#1} \right)}

\newcommand{\as}{{\alpha_s}}
\newcommand{\ee}{e^+e^-}
\newcommand{\ie}{i.e.\ }
\newcommand{\eg}{e.g.\ }

\newcommand{\caesar}{\textsc{caesar}}

\begin{document}

\title{Motivations and ideas behind hadron--hadron event shapes}
\author{G. Zanderighi}
  \affiliation{Physics Department,  Theory Division, CERN, 1211 Geneva 23,
    Switzerland}
\begin{abstract}
  We summarize the main motivations to study event shapes at hadron
  colliders. In addition we present classes of event shapes and show
  their complementary sensitivities to perturbative and
  non-perturbative effects, namely jet hadronization and underlying
  event.
\end{abstract}
\maketitle \thispagestyle{fancy}

\section{Introduction}
Event shapes describe the energy and momentum flow of the final state
in high energy collisions. Thanks to the synergy of their simplicity
and sensitivity to properties of QCD radiation, they are among the
most extensively studied observables in $e^+e^-$ and DIS collisions.
Specifically, studies of event shapes were used for measurements of
the strong coupling constant $\as$ and of its renormalization group
running~\cite{Bethke:2004uy}, for cross-checks of the gauge group
through measurements of the colour factors~\cite{SU3} and, most
important, they made it possible to get insights into the dynamics of
hadronization~\cite{Beneke,DasSalReview}.

Despite the great success of these studies in $e^+e^-$ and DIS, event
shapes have been largely neglected at hadron colliders, the only
(published) exceptions being a measurement of a variant of the
broadening by CDF in 1991~\cite{CDF-Broadening} and of a variant of
the thrust by D0 in 2002~\cite{D0Thrust}.
At hadron colliders, the counterpart of $\ee\to 2$ jets or DIS
[1+1]-jet event shapes are those in high-transverse momentum dijet
production.

Measurements and calculations are more difficult at hadron colliders.
Experimentally, one has to deal with the omnipresent underlying event
and with the impact of a limited detector coverage.  Theoretically,
the presence of four hard QCD-emitting partons at Born level implies
complicated patterns of interference between different dipoles, which
for the first time involve non-diagonal colour
structures~\cite{BottsSterman}.
Given these difficulties, it is natural to ask oneself whether there
is anything new that can be learned from event shapes at hadron
colliders. In the following I will address this question.

\section{New aspects of hadron--hadron event shapes}
It is worthwhile to set at first some terminology and conventions. A
given event shape is denoted by $V$, while $v$ denotes its value given
a set of momenta, and $L\equiv\ln(1/v)$ denotes its logarithm. By
convention, event shapes are defined in such a way that in the absence
of secondary emissions their value is precisely zero. A small value
then denotes the presence of only soft-collinear emissions. In this
region the smallness of the QCD coupling is compensated by large
logarithmic contributions. Fixed-order perturbative (PT) predictions
become unreliable and logarithmic enhanced terms have to be resummed
to all orders. The state-of-the-art is to resum terms up to
next-to-leading logarithms (NLL), \ie $\alpha_s^n L^{n}$, in the
exponent of the integrated distribution of $v$, $\Sigma(v) = \exp\{ L
g_1(\as L) + g_2(\as L) \}$~\cite{CTTW}.  Resummed predictions are
then matched to next-to-leading order (NLO) fixed-order calculations
(\ie $\cO{\as^2}$ relative to the Born process), so as to have an
accurate description in the whole phase space.

These accurate perturbative predictions do not account for
hadronization effects. Therefore, PT results are supplemented with
non-perturbative (NP) effects describing hadronization. Within the
dispersive approach~\cite{DMW}, one assumes that, while the
perturbative coupling diverges in the infrared (IR) because of the
presence of the Landau pole, the physical coupling is well defined and
integrable in the IR region.  One then introduces the average of the
full QCD coupling in the NP region
\begin{equation}
\alpha_0(\mu_I) \equiv \int_0^{\mu_I} \frac{d k_t}{\mu_I}  \as(k_t)\,. 
\label{eq:a0}
\end{equation}
In this framework it is possible to show that leading NP corrections
to event-shapes give rise to a shift $\delta V^{\rm NP}$ of the PT
distribution which turns out to be universal, \ie it is given in terms
of an observable-dependent, perturbatively calculable coefficient,
$c_v$, times the pure non-perturbative, observable-independent
parameter $\alpha_0$:
\begin{equation}
\Sigma(V) = \Sigma^{\rm PT}(V-\delta V^{\rm NP})\,,\qquad 
\delta V^{\rm NP}  = c_V \cdot \alpha_0\, \frac{ \mu_I}{Q}\,. 
\end{equation}
With today's techniques, $\alpha_0$ cannot be computed from first
principles, since this would require the knowledge of the QCD dynamics
in the IR, this information must therefore be extracted from data.
However, the beauty of this approach lies in its simplicity and
predictive power: having fitted $\alpha_0$ from one observable at a
given energy, one can predict leading $1/Q$ hadronization effects not
only at different energies, but also for different event shapes.

The dispersive approach turned out to be very successful
(see~\cite{DasSalReview} and references therein), in some cases even
so successful that theorists had to revisit their predictions in order
to agree with the experimental data~\cite{Dokshitzer:1998qp}.
It is however important to keep in mind that the dispersive approach
has been tested only with two-jet event shapes, \ie for those
observables for which the Born level event is made out of two hard QCD
partons and the first non-zero contribution starts with configurations
with three QCD partons.
While there exist theoretical calculations for non-perturbative
effects in the case of three-jet event shapes in $\ee$ or DIS
($[2+1]$) and they predict the same universal behaviour for the power
corrections~\cite{3jets} with the same parameter $\alpha_0$, this
important prediction has never been tested experimentally to date.

The calculation of power corrections to three-jet observables makes it
possible to test the dispersive approach in a new regime.  Indeed, for
two-jet event shapes it is sufficient to assume that NP radiation is
distributed uniformly in rapidity, as in the Feynman tube model, to
recover the predictions of the dispersive approach. For three-jet
event shapes, this is no longer true, since for radiation in between
jets there is simply no natural direction with respect to which one
can assume rapidity invariance of density of emitted hadrons.
Therefore a confirmation, or falsification, of such predictions would
shed light on the pattern of NP radiation.  For a more detailed
discussion, see~\cite{Banfiproc}.

A different NP approach is based on shape
functions~\cite{ShapeFunctions}, which account not only for leading NP
effects, but also for sub-leading ones ($\cO{1/(vQ)^n}$). Such an
approach is needed in the region of extremely small values of $v \sim
\Lambda_{\rm QCD}/Q$.  The price paid for the inclusion of
higher-order NP effects is the partial loss of predictive power: the
knowledge of the shape function for one observable is not sufficient
to predict hadronization effects for different observables. Two
important exceptions exist. One is represented by the class of
angularities in $e^+e^-$~\cite{Berger:2003pk}. This class of
observables is given in terms of a continuous parameter $-\infty< a< 2
$ ($a=0$ denotes the thrust and $a=1$ the broadening); and for them it
has been shown that shape functions are related by a simple scaling
rule~\cite{Berger:2003pk,Berger:2004xf}. This important theoretical
prediction has not yet been tested experimentally. A second class of
observables is the so-called fractional energy--energy moments (see
Appendix I.2 of~\cite{Banfi:2004yd}).  These observables have the same
NP scaling rule as the angularities, and, as angularities for $a<1$,
they are linear in the secondary emissions, which means that their
resummation is as simple as that of the thrust.

At hadron colliders, where two hard QCD partons are present already in
the initial state, any final-state measurement will lead beyond the
well-tested two-jet regime.
Additionally, unlike in $e^+e^-$ or DIS, in hadron--hadron collisions
multijet events are the natural playground for these QCD studies,
since they events appear at Born level, without any further
$\alpha_s$ suppression.
Both PT and NP effects are enhanced at hadron colliders, because of
the gluonic colour charges. For instance in $\ee$, at $\cO{\as}$
leading logarithms have an overall colour factor $2\,C_F$. In hadronic
dijets, for the pure gluonic channel, this is to be compared with
$4\,C_A$.
However, standard NP radiation is now always accompanied by the
presence of the underlying event. While this complicates the studies
of the NP corrections, if one is able to disentangle the two effects,
event shapes can provide useful information about the underlying
event. Currently, the only observable used to extract information
about the underlying event is the away-from-jet particle/energy
flow~\cite{Uevent}, which is however subject to larger theoretical
uncertainties than dijet event shapes.
In the following I will describe how event shapes can be constructed
which deliberately enhance or suppress the effect of the underlying
event, thereby limiting its impact for purely perturbative studies, or
enhancing it in order to focus on non-perturbative effects.

\section{Observables}
The way resummed predictions are currently obtained in hadron--hadron
collisions is with the automated resummation tool
\caesar~\cite{Banfi:2004yd}. A current limitation of \caesar\, (and of
analytical resummations) is that the observable should be
global~\cite{NG1}, \ie sensitive to emissions everywhere in phase
space. This is because resummations are based on the independent
emissions approximation~\cite{CTTW}, which is known to fail to
describe all NLL for non-global observables, because of configurations
with large-angle soft gluons in the unobserved region, which emit
large-angle energy ordered gluons in the observed region.
At hadron colliders this becomes a serious drawback, since detectors
can cover only a limited rapidity region. Notably no measurements can
be performed too close to the beam. This is usually parameterized with
a maximum accessible rapidity $\eta_0$.

Theoretical and experimental requirements thus seem to be in conflict.
In spite of this, it was shown in~\cite{Banfi:2004nk} that three
different classes of observables can be designed to overcome this
tension. In the following I will introduce these observables and show
that they have complementary sensitivities, specifically with respect
to the underlying event and jet hadronization effects.

\subsection{Directly global observables}
\begin{figure}
\begin{minipage}{.48\textwidth}
\begin{flushleft}
\hspace{1.5cm}$\displaystyle \blue y_{23} \sim 1 $
\end{flushleft}
\begin{center}
\includegraphics[width=.8\textwidth]{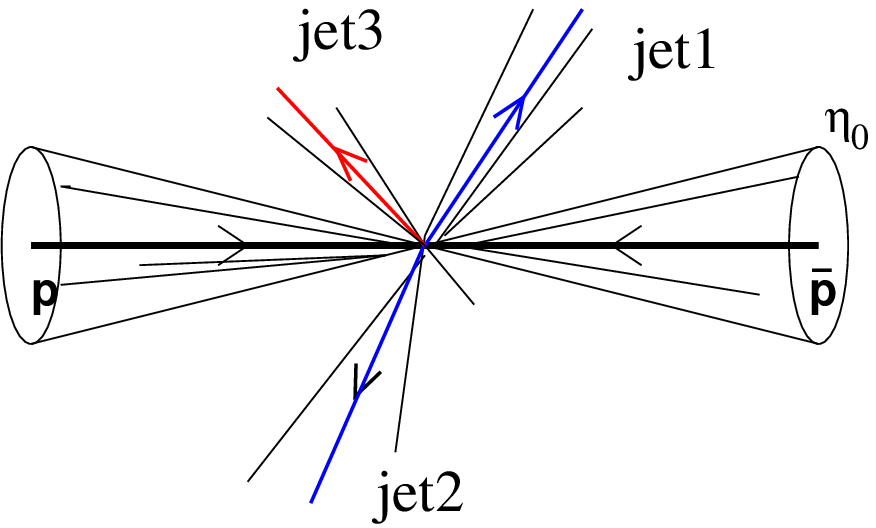}
\end{center}
\end{minipage}
\begin{minipage}{.48\textwidth}
\begin{flushleft}
\hspace{1cm}$\displaystyle \blue y_{23} \ll 1 $
\end{flushleft}
\begin{center}
\includegraphics[width=.8\textwidth]{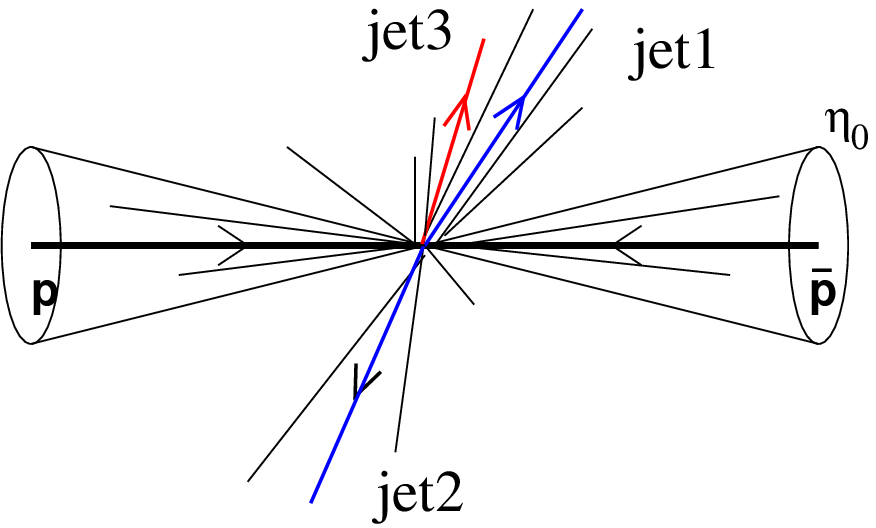}
\end{center}
\end{minipage}
\caption{Illustration of a three-jet event, $y_{23} \sim1$ (left) and of a two-jet-like event, $y_{23} \ll 1$ (right)}
\label{fig:y3}
\end{figure}

To define directly global observables one selects events with two
large transverse momentum jets and defines observables similar to
$e^+e^-$, but purely in the transverse plane. For instance the directly
global transverse thrust is given by
\begin{equation}
  T_{\perp, g} \equiv \max_{\vec{n}_\perp} \frac{ \sum_i
      |\vec{p}_{\perp i}\cdot {\vec{n}_\perp |}}{\sum_i 
|\vec{p}_{\perp i}|}\,,\qquad 
\tau_{\perp, g} \equiv 1- T_{\perp, g}\,, 
\label{eq:othrglob}
\end{equation}
where $\vec{p}_{\perp, i}$ denotes the transverse momentum with
respect to the beam of parton $i$ and the directly global thrust minor
is defined as
\begin{equation}
T_{m, g} \equiv \frac{ \sum_i |\vec{p}_{\perp i}\times 
  {\vec{n}_{\perp} |}}{\sum_i |\vec{p}_{\perp i}|}\,. 
\label{eq:tmglob}
\end{equation}
Similarly to event shapes, one can define three-jet resolution
variables. In the longitudinal invariant exclusive $k_t$ algorithm one
defines the distance between particle $i$ and the beam $B$ and between
two particles $i,j$
\begin{equation}
d_{iB} \equiv q_{\perp i}^2\,,
\qquad 
d_{ij} \equiv \min\{q_{\perp i}^2, q_{\perp j}^2
\}\left((\eta_i-\eta_j^2)+(\phi_i-\phi_j)^2\right)  
\end{equation}
and then recombines the pair with smallest distance. Finally, one
defines
\begin{equation}
y_{23} \equiv \frac{1}{(E_{\perp 1}+E_{\perp 2})^2}\max_{n \ge 3} d^{(n)}\,,
\label{eq:y23glob}
\end{equation}
where $d^{(n)}$ is the smallest distance when $n$ pseudo-particles are
left over after recombination and $E_{\perp 1,2}$ denote the
transverse energies of the two leading jets. As for dijet event
shapes, $y_{23} \ll 1$ denotes a two-jet-like configuration (see
Fig.~\ref{fig:y3}).

For directly global observables, measurements should be carried out as
forward as possible (up to $\eta_{0} \sim 3.5$ at the Tevatron and
$\eta_{0} \sim 5$ at the LHC).
It has been shown~\cite{KoutZ0} that measurements in the unobserved
region are negligible at NLL as long as the observable is not too
small.  More precisely, given an observable which, in the presence of a
single soft emission collinear to leg $\ell$, behaves as
\begin{equation}
  \label{eq:vpar}
 V \sim d_\ell \left(\frac{k_t}{Q}\right)^{a} e^{-b_\ell
    \eta} g_\ell(\phi)\,, 
\end{equation}
the formal limit for NLL predictions to be valid reads 
\begin{equation}
  \label{eq:vlim}
\ln (X\cdot V)  \gtrsim -(a +b)\cdot
\eta_0\,, \qquad b \equiv \min\{b_1,b_2\}\,. 
\end{equation}
Here $X$ denotes a rescaling of the argument of the logarithm chosen
so as to cancel the average value of $d_\ell g_\ell(\phi)$ in the
resummed exponent $g_2(\as L)$,
see~\cite{Dasgupta:2002dc,Banfi:2004nk} and 1,2 label the incoming
legs.
The strategy is then to neglect $\eta_{0}$ in theoretical
predictions and to check a posteriori which portion of the tail is
beyond control.

\subsection{Recoil enhanced observables}
\begin{figure}[b]

  \includegraphics[width=.3\textwidth]{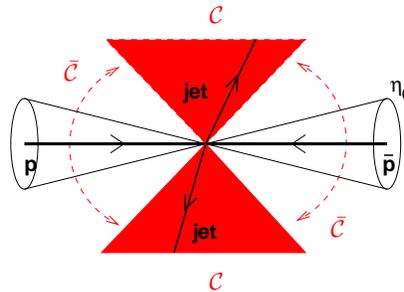}
  \psfrag{C}{\tiny $\red \cC$} \psfrag{B}{\tiny $\red \barcC$}
\caption{The central region $\cC$, containing the jets, and the forward rapidity cut $\eta_0$.}
\label{fig:cc}
\begin{picture}(1,1)(0,0)%
\put(0,24){\begin{pspicture}(0,0)(2,2)
$\red \cC$ 
\end{pspicture}}
\put(0,125){\begin{pspicture}(0,0)(2,2)
$\red \cC $ 
\end{pspicture}}
\put(-55,103){\begin{pspicture}(0,0)(2,2)
$\red \barcC$ 
\end{pspicture}}
\put(46,44){\begin{pspicture}(0,0)(2,2)
$\red \barcC$ 
\end{pspicture}}
\end{picture}
\end{figure}

The definition of recoil enhanced observables proceeds as follows:
define a central region $\cal C$, \eg $|\eta| < \eta_{\rm max} \sim
\cO{1}$, which should contain the two jets, see Fig.~\ref{fig:cc}.
Then define central observables using particles only in the central
region. For instance, the central transverse thrust is given by
\begin{equation}
\label{eq:cenobs}
  T_{\perp,  \cal C} \equiv \max_{\vec{n}_\perp, \cC} \frac{ \sum_{ i\in \cal C}
      |\vec{p}_{\perp i}\times {\vec{n}_{\perp, \cC} |}}{\sum_{ i\in \cal C} |\vec{p}_{\perp i}|}\,, \qquad 
\tau_{\perp,  \cal C} \equiv 1- T_{\perp,  \cal C}\,. 
\end{equation}
Similarly one defines a central thrust minor, or three-jet resolution
variable, as in eqs.~(\ref{eq:tmglob}) and~(\ref{eq:y23glob}) but
restricting the measurement to the central region.

Transverse-momentum conservation ensures that
\begin{equation}
\label{eq:rperpqperp}
  \cR_{\perp,  \cC} \equiv \frac{1}{Q_{\perp, \cC}} \Bigg|\sum_{ i \in \cC} {\vec
        q}_{\perp i}\Bigg| 
{=} \frac{1}{Q_{\perp, \cC}} \Bigg|\sum_{ i \not \in \cC} {\vec
        q}_{\perp i}\Bigg|\,, 
\qquad 
    Q_{\perp, \cC} \equiv \sum_{ i\in \cC} q_{\perp i}\,. 
\end{equation}
Central observables can then be made global by adding the recoil term
$\cal R_{\perp, \cal C} $ (or a power of it), \eg the recoil enhanced
transverse thrust, the thrust minor and the three-jet resolution
parameter read
\begin{equation}
\tau_{\perp, \cal R} \equiv \tau_{\perp, \cal C} + {\cal R_{\perp, {\cal C}}}
\>, \qquad
T_{m, \cal R} \equiv T_{m, \cal C} + {\cal R_{\perp, \cal C}}\>,\qquad
y_{23, \cal R} \equiv y_{23, \cal C} + {\cal R}^2_{\perp, \cal C}\>.
\label{eq:tautmy3rec}
\end{equation}
Note that to ensure continuous globalness~\cite{Dasgupta:2002dc}
(roughly speaking that the transverse-momentum dependence be
independent of the emission's direction) one has to add a second power
of the recoil term to $y_{23, \cal C}$.

Observables in this class are global despite the fact that
measurements are truly restricted to a central region. This is
because they have an indirect sensitivity to forward emissions through
recoil effects. This mechanism was already well known from some DIS
observables~\cite{Dasgupta:2002dc}.

\subsection{Observables with exponentially suppressed forward terms}
The last class of observables discussed is the one with exponentially
suppressed forward terms. Also in this case one starts defining
central observables, as in eq.~(\ref{eq:cenobs}).
One then introduces the mean transverse-energy weighted rapidity
$\eta_\cC$ of the central region 
\begin{equation} 
  \eta_\cC = \frac{1}{
    Q_{\perp,\cC}} \sum_{i\in\cC} \eta_i\, q_{\perp i}\,,\qquad\quad
\end{equation}
and defines an exponentially suppressed forward term as follows
\begin{equation} 
\cE_{\barcC} \equiv \frac{1 }{Q_{\perp,\cC}}
  \sum_{i \notin \cC} q_{\perp i} \,e^{-|\eta_i - {\eta_\cC}|}\,,  
\end{equation}
where the total transverse momentum $Q_{\perp,_\cC}$ of the central
region is defined in eq.~\eqref{eq:rperpqperp}. 
With these elements, one can make the usual central observables
global by adding the exponentially suppressed forward term
$\cE_{\barcC}$ (or a power of it to ensure continuous globalness).
For instance, the exponentially suppressed transverse thrust, thrust
minor and three-jet resolution parameter are given by
\begin{equation}
\tau_{\perp, \cal E} \equiv \tau_{\perp, \cal C} + {\cE}_\barcC \>,\qquad 
T_{m, \cal E} \equiv T_{m, \cal C} + {\cE}_\barcC \>,\qquad 
y_{23, \cal E} \equiv y_{23, \cal C} + {\cE}^2_\barcC\>. 
\end{equation}

Additionally, one can consider observables that are more suitably
defined only in a restricted (central) region.
One first separates the central region $\cC$ into an up part $\cC_U$, 
consisting of all particles in $\cC$ with $\vec p_{\perp} \cdot \vec
n_{T,\cC} > 0$, and a down part $\cC_D$, consisting of all particles in
$\cC$ with $\vec p_{\perp} \cdot \vec n_{T,\cC} < 0$.

One then defines, in analogy to $\ee$~\cite{Clavelli}, the normalized
squared invariant masses of the two regions:
\begin{equation}
  \label{eq:mass-XC}
  \rho_{X,\cC} \equiv \frac{1}{Q_{\perp,\cC}^2}
  \left(\sum_{i\in \cC_X} q_{i}\right)^2\,,\qquad X = U, D\,,
\end{equation}
from which one can obtain a (non-global) central sum of masses and
a (non-global) heavy-mass:
\begin{equation}
  \label{eq:mass-C-sum-heavy}
  \rho_{S,\cC} \equiv \rho_{U,\cC} + \rho_{D,\cC}\,,\qquad\quad
  \rho_{H,\cC} \equiv \max\{\rho_{U,\cC}, \rho_{D,\cC}\}\,,
\end{equation}
together with versions that include the addition of the
exponentially suppressed forward term:
\begin{equation}
  \label{eq:mass-E-sum-heavy}
  \rho_{S,\cE} \equiv \rho_{S,\cC} + \cE_{\barcC}\,,\qquad\quad
  \rho_{H,\cE} \equiv \rho_{H,\cC} + \cE_{\barcC}\,.
\end{equation}

This separation into up and down regions can also be used to define
boost-invariant jet broadenings. One first introduces rapidities and
azimuthal angles of axes for the up and down regions:
\begin{equation}
  \label{eq:broadening-axes}
  \eta_{X,\cC} \equiv \frac{\sum_{i\in \cC_X} q_{\perp i}
    \eta_i}{\sum_{i\in \cC_X} 
    q_{\perp i}}\,,\qquad
  \phi_{X,\cC} \equiv \frac{\sum_{i\in \cC_X} q_{\perp i}
    \phi_i}{\sum_{i\in \cC_X} 
    q_{\perp i}}\,,\qquad
   X = U, D\,,
\end{equation}
and defines broadenings for the two regions:
\begin{equation}
  \label{eq:BX-C}
  B_{X,\cC} \equiv \frac{1}{2Q_{\perp,\cC}} \sum_{i\in\cC_X}
  q_{\perp i}\sqrt{(\eta_i-\eta_{X,\cC})^2 + (\phi_i - \phi_{X,\cC})^2}\,,
  \qquad X = U, D\,. 
\end{equation}
The central total and wide-jet broadenings are then given by 
\begin{equation}
  \label{eq:B-C-total-wide}
  B_{T,\cC} \equiv B_{U,\cC} + B_{D,\cC}\,,\qquad\quad
  B_{W,\cC} \equiv \max\{B_{U,\cC}, B_{D,\cC}\}\,.
\end{equation}
Exponentially suppressed, global observables are then obtained by
adding the forward term $\cE_{\barcC}$
\begin{equation}
  \label{eq:BTWE}
  B_{T,\cE} \equiv B_{T,\cC} + \cE_{\barcC}\,,\qquad
  B_{W,\cE} \equiv B_{W,\cC} + \cE_{\barcC}\,.
\end{equation}

\section{Sample NLL distributions from \caesar}
To illustrate different features of resummed predictions, we present
here some sample distributions as obtained with \caesar.
We select dijet events at the Tevatron Run II regime, $\sqrt{s} =
1.96$ TeV, by running the longitudinally invariant $k_t$
algorithm~\cite{JetratesHH-CDSW} and by requiring the presence of two
jets in the central region ($|\eta| < 0.7$), the transverse momentum
of the hardest jet satisfying $E_\perp > \mbox{E}_{\perp, \rm min} =
50\mbox{ GeV}$ (unless differently specified). We define the central
region $\cal C$ $|\eta| < \eta_{\rm max } = 1.1$. This ensures that
the two jets are well contained in $\cal C$.

Most of the properties of distributions can be understood in terms of
the coefficients $a$ and $b_\ell$ appearing in the parametrization of
the observable given in eq.~(\ref{eq:vpar}). This is because the
leading Sudakov effect is given by
\begin{equation}
\Sigma(V) \sim e^{\frac{\as}{2\pi}G_{12} L^2+\dots}\,,\qquad \qquad 
G_{12} = -\frac{2}{a}\sum_\ell \frac{C_\ell}{a+b_\ell}\,, 
\end{equation}
so that it is determined only by the colour charges $C_\ell$ ($C_\ell=
C_F, C_A$ if parton $\ell$ is a quark/gluon) and by the values of $a$
and $b_\ell$.
The point up to which the resummation is under NLL control, given in
eq.~(\ref{eq:vlim}), also depends on the values of $a$ and $b_\ell$
(although it also depends on the values of $d_{\ell} g_\ell(\phi)$
through $X$).

\begin{table}[t]
\begin{center}
\begin{minipage}{0.4\textwidth}
\begin{tabular}{|c||c|c||c|c|}
\hline
 $V$ &  $a_{\ell}$ (inc) &  $b_{\ell}$ (inc) &  $a_{\ell}$ (out) &  $b_{\ell}$ (out) \\
\hline
\hline
$\tau_{\perp, g}$ & 1 & 0 & 1& 1 \\
\hline
 $T_{m, g}$ & 1 & 0 & 1 &  0 \\
\hline
 $y_{23,g}$ &  2 &  0 &  2 &  0 \\
\hline
\end{tabular}
\end{minipage}
\begin{minipage}{0.4\textwidth}
\begin{tabular}{|c||c|c||c|c|}
\hline
 $V$ &  $a_{\ell}$ (inc) &  $b_{\ell}$ (inc) &  $a_{\ell}$ (out) &  $b_{\ell}$ (out) \\
\hline
\hline
$\rho_{H, \cE}$ & 1 & 1 & 1& 1 \\
\hline
$B_{W, \cE}$ & 1 & 1 &1 &0 \\
\hline
\end{tabular}
\end{minipage}
\end{center}
\caption{Coefficients $a$ and $b_\ell$ for some global (left) and exponentially suppressed (right) observables as they appear in eq.~(\ref{eq:vpar}), given for incoming and outgoing legs. 
}
\label{tab:vcoeffs}
\end{table}

In Table~\ref{tab:vcoeffs} we report the values of the coefficients $a$
and $b_\ell$ for various observables as computed by \caesar\, (despite
the presence of four legs it is sufficient to distinguish only between
in/outgoing legs).

\begin{figure}[b]
\begin{minipage}{.3\textwidth}
  \includegraphics[width=0.95\textwidth]{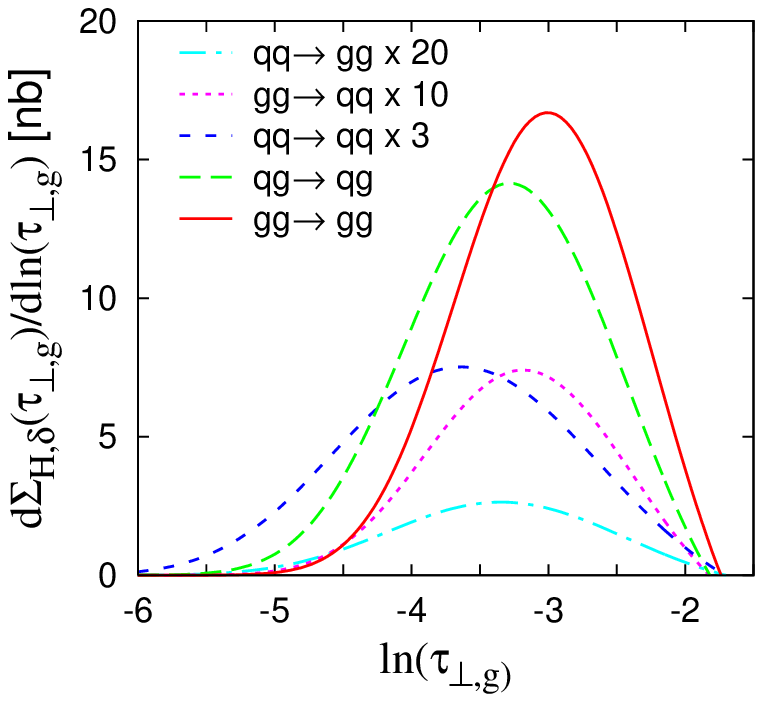}
  \end{minipage}
\begin{minipage}{.3\textwidth}
\includegraphics[width=0.95\textwidth]{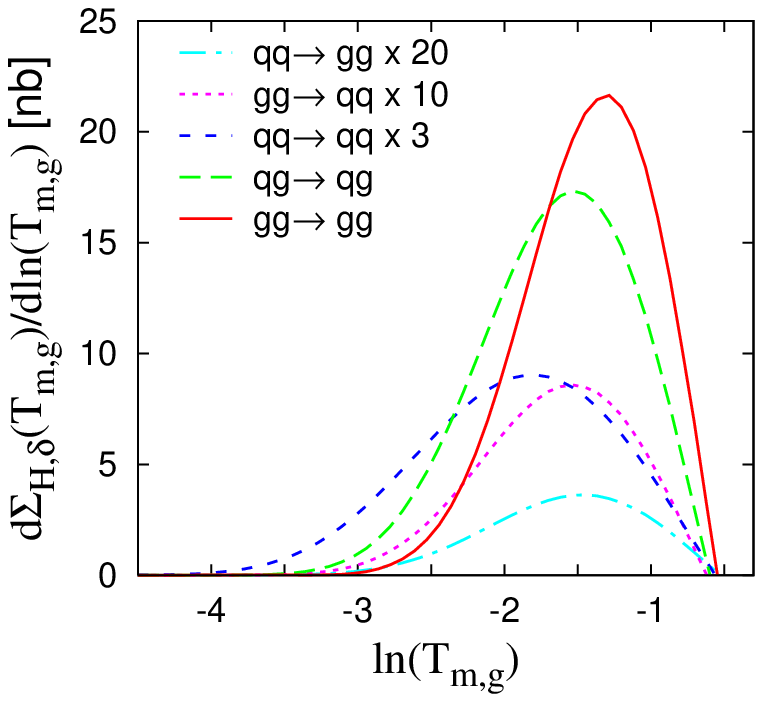}
  \end{minipage}
\begin{minipage}{.3\textwidth}
\includegraphics[width=0.95\textwidth]{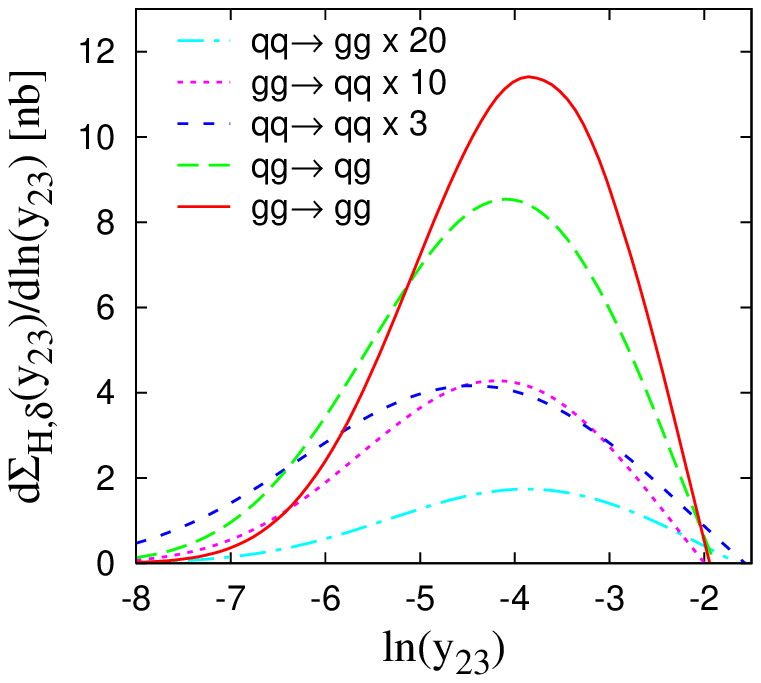}
  \end{minipage}
\caption{NLL resummed differential
  distributions for the directly global thrust (left), thrust minor
  (central) and three-jet resolution parameter (right), for various
  partonic channels.}
\label{fig:vglob}
\begin{picture}(1,1)(0,0)%
\put(-203,50){\begin{pspicture}(0,0)(2,2)
\psframe[linecolor=black,linestyle=none,hatchcolor=black,fillstyle=vlines,hatchangle=-45](0.7,1.120)(1.0,2.5)
\psline[linecolor=black,linestyle=solid](1.0,1.120)(1.0,2.5)
\end{pspicture}
}
\put(-48,50){\begin{pspicture}(0,0)(2,2)
\psframe[linecolor=black,linestyle=none,hatchcolor=black,fillstyle=vlines,hatchangle=-45](0.7,1.120)(1.0,2.5)
\psline[linecolor=black,linestyle=solid](1.0,1.120)(1.0,2.5)
\end{pspicture}
}
\put(97,50){\begin{pspicture}(0,0)(2,2)
\psframe[linecolor=black,linestyle=none,hatchcolor=black,fillstyle=vlines,hatchangle=-45](0.7,1.120)(1.0,2.5)
\psline[linecolor=black,linestyle=solid](1.0,1.120)(1.0,2.5)
\end{pspicture}
}
\end{picture}  
\end{figure}

In Fig.~\ref{fig:vglob} we show pure NLL resummed differential
distributions for various global observables, the thrust, thrust
minor, and three-jet resolution parameter, defined in
eqs.~(\ref{eq:othrglob}),~(\ref{eq:tmglob}), and~(\ref{eq:y23glob})
for various partonic channels. The figure shows that the thrust minor
is peaked at higher values of $v$, while the peaks of $y_{23}$ lie at
lowest values.  This is expected since for these observables the
position of the peak depends only on the coefficients of the total
colour charge of in/outgoing partons, $C_{{\rm in}/{\rm out}}$
(specifically here $G_{12} = -(2C_{\rm in} + C_{\rm out}), -2(C_{\rm
  in} + C_{\rm out}), -(C_{\rm in} + C_{\rm out})$ for the thrust,
thrust minor, and three-jet resolution, respectively).  The figure
also shows that tails of quark-channel distributions generally extend
to lower values of $v$ compared to channels dominated by gluons. This
can again be inferred from the value of $G_{12}$ for the given
channel; it simply reflects the fact that quarks radiate less then
gluons.
The region to the left of the thick line is
beyond control of the NLL resummed predictions, given in
eq.~(\ref{eq:vlim}). Note that $X\sim 5$ for the thrust, while $X=1$
for the thrust minor and $y_{23}$, so that for these last two
observables the cut is located simply at $\eta_0$ and $2\eta_0$
respectively. We see that for the thrust a small part of the tail
extends beyond the cut, while for the other two observables almost all
the distributions are under perturbative control.

Finally, one should remark that at higher values of $v$ NLL
distributions become negative. This unphysical behaviour will be
corrected by the fixed-order predictions once the matching with fixed
order results is carried out.

\begin{figure}
\begin{minipage}{.3\textwidth}

  \includegraphics[width=0.95\textwidth]{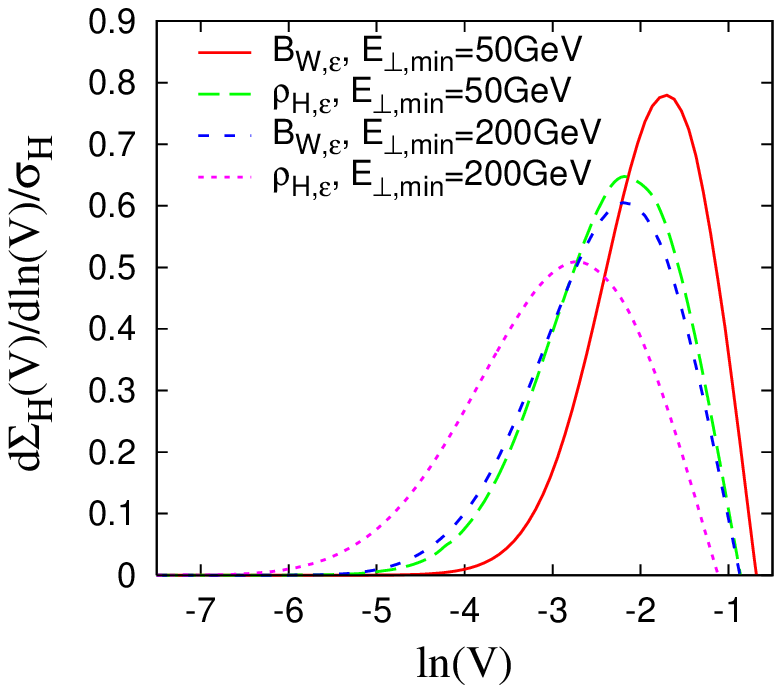}
  \end{minipage}
\begin{minipage}{.3\textwidth}
\includegraphics[width=0.95\textwidth]{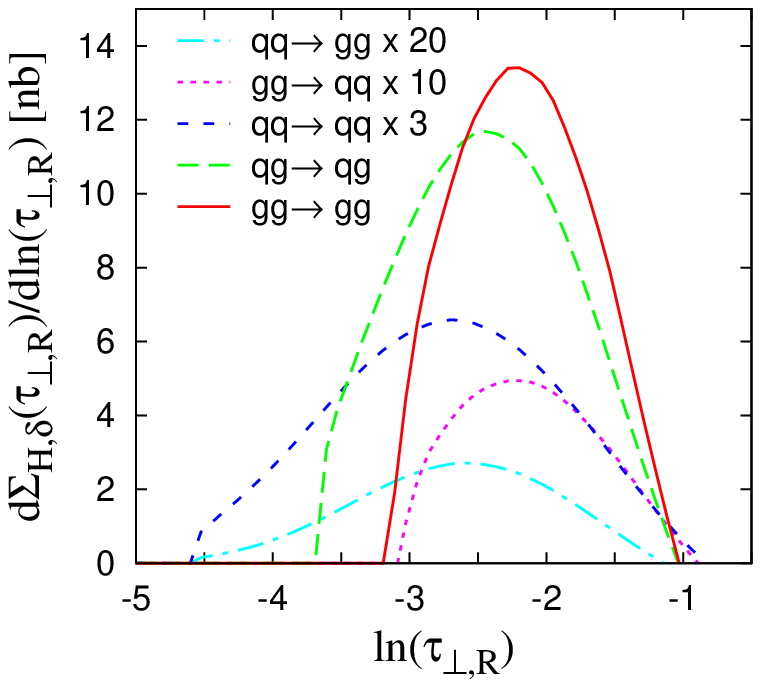}
  \end{minipage}
 \begin{minipage}{.3\textwidth}
\includegraphics[width=0.95\textwidth]{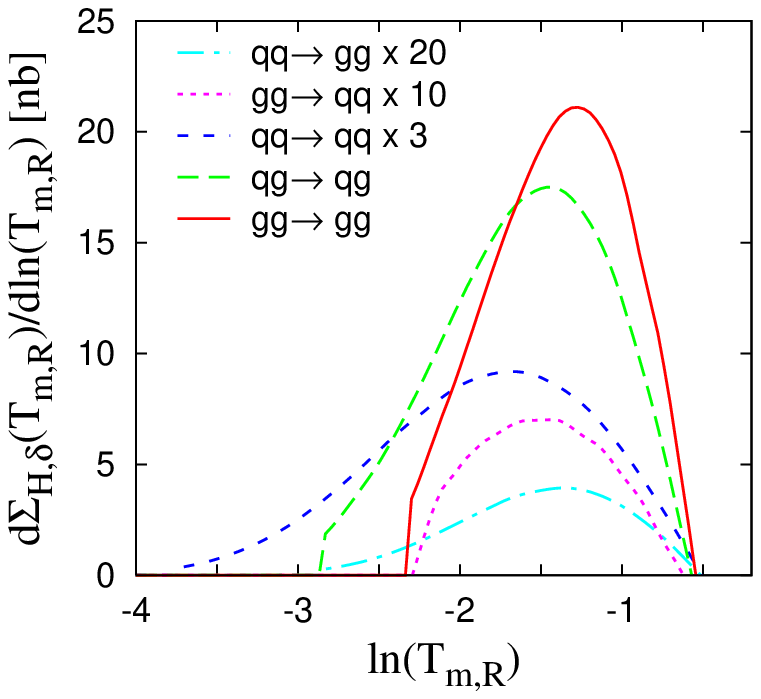}
  \end{minipage}
\caption{Total differential
  distribution for exponentially suppressed wide broadening and heavy
  jet-mass for two valued of $E_{\perp, \rm min}$ (left).
  Differential distributions for the recoil enhanced thrust (central)
  and thrust minor (right) for various partonic channels.}
\label{fig:vexprec}
\begin{picture}(1,1)(0,0)%
\put(-218,52){\begin{pspicture}(0,0)(2,2)
\psframe[linecolor=black,linestyle=none,hatchcolor=black,fillstyle=vlines,hatchangle=-45](0.7,1.120)(1.0,2.5)
\psline[linecolor=black,linestyle=solid](1.0,1.120)(1.0,2.5)
\end{pspicture}
}
\end{picture}  
\end{figure}

In Fig.~\ref{fig:vexprec} (left) we show total distributions, summed
over all partonic channels, for two exponentially suppressed
observables, a variant of the heavy jet-mass, and the wide broadening
defined in eqs.~(\ref{eq:mass-E-sum-heavy}) and~(\ref{eq:BTWE}) for
two values of $E_{\perp, \rm min}$. The two observables have a
different leading Sudakov suppression ($G_{12} = - (C_{\rm in}+2
C_{\rm out}),- (C_{\rm in}+C_{\rm out})$ for the broadening and the
jet mass respectively).
Compared to directly global observables, one can see a remarkable
improvement as far as the position of the cut is concerned, due to the
exponentially suppressed term ($b_\ell = 1$ for incoming legs).

One can also see the effect of changing the transverse energy cut:
increasing the cut, the distribution is peaked at lower values.  This
is due to two different mechanisms: with higher transverse energy cuts
one probes the coupling at higher scales and the parton densities at
higher $x$, where quark channels become more important.

Finally, in Fig.~\ref{fig:vexprec} (centre and right) we show the
distribution for the recoil-enhanced transverse thrust and thrust
minor defined in eq.~(\ref{eq:tautmy3rec}). Since these observable
measure only very central emissions, the $\eta_0$ cut has no effect.
However, the distributions have a different cut since the resummation
breaks down at some value of $v$ (which generally depends on the
channels too).
This breakdown is well understood~\cite{DSBroad,RakowWebber}, and it
is a characteristic of all those observables for which contributions
from multiple emissions can cancel one another. What happens is that,
for sufficiently small $v$, it becomes more likely to suppress the
value of $v$ via cancellations between emissions rather than by
Sudakov suppression. This leading logarithmic (LL) change of behaviour
cannot be accounted for by the NLL resummation.
Points where the NLL resummation is not reliable have been excluded
from the plots. The effect of this can be seen in the cut of the tails
of the distributions.

\section{Complementarity between the observables}

The above observables have complementary properties, sensitivities and
advantages/disadvantages, which makes them a very rich field of
activity.
To be more precise, directly global observables are simple, being
defined analogously to $\ee$ observables, just purely in the
transverse plane.  However, the impact of $\eta_0$ is not always
negligible and this can be checked only a posteriori.
Observables with recoil term measure only central emissions. The
recoil term can however be subject to large cancellations, making an
accurate determination of it problematic. Additionally, the Sudakov
effect competes with a cancellation in the vectorial sum in the recoil
term, causing the resummation to break down at some point.
Observables with an exponentially suppressed forward term have a very
low sensitivity to the forward region; they are thus the most
promising compromise. Yet experimental measurements are complicated by
the fact that one needs to merge measurements from different detectors
(tracking information in the central region, where a good resolution
is needed, calorimeter information in the forward region, where an
``average'' information on the momentum flow is sufficient).

These theoretical complementarities are summarized in the first two
columns of Table~\ref{tab:summary} for variants of the thrust
($\tau_\perp$), thrust-minor ($T_m$), broadenings ($B_X$), jet-masses
($\rho_X$), and three-jet resolution parameter ($y_{23}$).
Additionally the table shows the sensitivity of these observables to
the underlying event and the hadronization of the jets.
In the following we will explain the listed properties. 

\newcommand\TT{\rule{0pt}{3.3ex}}
\newcommand\BB{\rule[-2.2ex]{0pt}{0pt}}
\begin{table}[t]
\begin{center}
  \begin{tabular}[t]{|c|c|c|c|c|} \hline
    Event shape & Impact of $\eta_{0}$ & \TT\BB
    \begin{minipage}[c]{0.18\textwidth}
      \begin{center}
        Resummation breakdown
      \end{center}
    \end{minipage}
    & 
    \begin{minipage}[c]{0.15\textwidth}
      \begin{center}
        Underlying event
      \end{center}
    \end{minipage}
    & 
    \begin{minipage}[c]{0.17\textwidth}
      \begin{center}
        Jet hadronization
      \end{center}
    \end{minipage}\\ \hline\hline
    $\tau_{\perp,g}$ & tolerable$^{*}$ 
& none & $\sim \eta_{0}/Q$ & $\sim 1/Q$
    \\ \hline
    $T_{m,g}$ & tolerable & none& $\sim \eta_{0}/Q$ & $\sim 1/(\sqrt{\as} Q)$
    \\ \hline
    $y_{23}$ & tolerable & none& $\sim \sqrt{y_{23}}/Q^*$ & $\sim \sqrt{y_{23}}/Q^*$
    \\ \hline\hline
    $\tau_{\perp,\cE}$, $\rho_{X,\cE}$ & negligible & none& $\sim 1/Q$ & $\sim 1/Q$
    \\ \hline
    $B_{X,\cE}$ & negligible & none& $\sim 1/Q$ & $\sim
    1/(\sqrt{\as} Q)$
    \\ \hline
    $T_{m,\cE}$ & negligible & serious& $\sim 1/Q$ & $\sim
    1/(\sqrt{\as} Q)$
    \\ \hline
    $y_{23,\cE}$ & negligible & none& $\sim 1/Q$ & $\sim \sqrt{y_{23}}/Q^{*}$
    \\ \hline\hline
    $\tau_{\perp,\cR}$, $\rho_{X,\cR}$ & none & serious & $\sim 1/Q$ & $\sim 1/Q$
    \\ \hline
    $T_{m,\cR}$, $B_{X,\cR}$ & none & tolerable  & $\sim 1/Q$ & $\sim
    1/(\sqrt{\as} Q)$
    \\ \hline
    $y_{23,\cR}$ & none & intermediate$^{*}$ &  $\sim \sqrt{y_{23}}/Q^*$ & $\sim \sqrt{y_{23}}/Q^*$
    \\ \hline
  \end{tabular}
\end{center}
\caption{Summary of the complementary properties of various observables. Entries marked with a * are subject to uncertainty at present.}
\label{tab:summary}
\end{table}

Let us begin by considering effects of jet hadronization.  Studies of
$\ee$ observables showed that, for those that are exponentially
suppressed in rapidity, the power corrections amount to a rigid shift
$\propto 1/Q$~\cite{Dokshitzer:1997iz}, while those that are uniform
in rapidity power corrections in addition lead to a squeeze of the
distribution which is $\propto
1/(\sqrt{\as}Q)$~\cite{Dokshitzer:1998qp}:
\begin{displaymath}
\begin{array}{lllllll}
  V & \sim & \frac{k_t}{Q}e^{-|\eta|} &\Rightarrow & 
\delta V^{\rm NP} & \sim &  c_V \alpha_0\,  \frac{\mu_I}{Q}, \\
  V & \sim & \frac{k_t}{Q} & \Rightarrow & \delta V^{\rm NP} & 
\sim & c_V(v) \alpha_0\, \frac{\mu_I}{\sqrt{\as}Q}.  \\
\end{array}
\end{displaymath}
The origin of the observed difference is the following. For an
observable $V$, which behaves as $V \sim \frac{k_t}{Q} f_V(\eta)$,
for a fixed perturbative configuration, the contribution to the power
correction from parton with colour charge $C_\ell$ can be written
schematically as
\begin{equation}
\delta V  \sim \int_{\rm NP} \frac{dk_\perp}{k_\perp} d\eta 
\frac{C_\ell \alpha_s(k_\perp)}{\pi} \frac{k_\perp}{Q}\, f_V(\eta)\,,  
\end{equation}
where the integration is over the region where the gluon is
non-perturbative.

For observables which are exponentially damped in rapidity, the
rapidity integration can be extended up to infinity. The rapidity and
transverse-momentum dependence then completely factorize: the rapidity
part goes into the observable dependence coefficient $c_V = \int d\eta
f_V(\eta)$, while the transverse-momentum integral gives rise to
$\alpha_0$, the average of the coupling in the IR region, below some
merging scale $\mu_I$, given in eq.~(\ref{eq:a0}). One obtains
\begin{equation}
\delta V  \propto c_V \alpha_0\, \frac{\mu_I}{Q}\,. 
\end{equation}

For observables which are uniform in rapidity, the rapidity integral
cannot be extended up to infinity and PT radiation provides an
effective cut-off in the rapidity integration. This cut-off is
generally given by the angle between the direction of hard emissions
and the reference directions used for the measurements (\eg thrust
axis in 2-jet events, event-plane for near-to-planar 3-jet events).
This angle depends logarithmically on the transverse momentum of the
hard emitting parton $q_t$.  Averaging this contribution over the
Sudakov PT recoil distribution $e^{-\as \ln^2(q_t/Q)}$ amounts to a
singular correction $\propto 1/\sqrt{\as}$.

This same pattern is expected to hold in hadron--hadron collisions as
well.
The parametrization of the observable given in eq.~(\ref{eq:vpar}),
which is established automatically by \caesar, allows one to determine
the rapidity behaviour of the observable (\ie the value of the
$b_\ell$) and therefore the type of power correction. Specifically,
observables with $b_{\ell} = 1$ for outgoing legs will have power
corrections $1/Q$, while those with $b_{\ell} = 0$ for outgoing legs
will have power corrections $1/(\sqrt{\as}Q)$. Finally, one sees from
the table that power corrections for the $y_{23}$ jet-resolution
parameters are estimated to be $\propto \sqrt{y_{23}}/Q$. The reason
is that, contrary to the other observables, which are linear in $k_t$,
$y_{23}$ is quadratic, therefore the effect of the power correction
can be estimated via
\begin{eqnarray}
\delta y_{23} &= &y_{23, \rm NP} - y_{23, \rm PT}
\propto \int_{0}^{\mu_I} \frac{d k_{\perp, \rm NP}}{Q} d\eta\, \as(k_\perp) 
\left( \frac{(k_{\perp, \rm PT} + k_{\perp, \rm NP})^2}{Q^2} 
- \frac{k_{\perp, \rm PT}^2}{Q^2}
\right)\\ 
&\sim &\int_{0}^{\mu_I} \frac{d k_{\perp}^{\rm NP}}{Q} \as(k_\perp) 
\frac{k_\perp^{\rm PT} k_\perp^{\rm NP}}{Q^2} 
\sim \sqrt{y_{23}}\alpha_0 \frac{\mu_I }{Q} \,. 
\end{eqnarray}

Finally, let us motivate how the underlying event affects measurements
of these observables.
We assume that particles from the underlying event are distributed
uniformly in rapidity~\cite{tube-model}; therefore the effect of the
underlying event to an observable $V$, parametrized according to
eq.~(\ref{eq:vpar}), can be estimated by
 \begin{equation}
  \label{eq:tautg-underlying-eventint}
  \langle \delta^{(\rm U.E.)}  V\rangle \sim 
    \int_{{ -\eta_0}}^{ \eta_{0}} d\eta 
  \int dk_\perp \frac{d\phi}{2\pi} \frac{dn^{(\rm U.E.)}}{d\eta dk_\perp}
  d_{1,2} \left(\frac{k_\perp}{Q}\right)^a e^{-b_{1/2}|\eta|} g_{1,2}(\phi)
\sim  \frac{\langle k_\perp^a\rangle^{(\rm U.E.)}}{Q^a} \langle g_{1,2}(\phi)\rangle 
\int_{{ -\eta_0}}^{ \eta_{0}}d\eta\, 
e^{-b_{1,2} |\eta|}
    \,.   
\end{equation}
If we consider observables for which $b_{\ell} = 0$ for incoming legs,
the underlying event will be enhanced by $\sim \eta_0$, while for
observables with $b_{\ell} = 1$, the rapidity integral can be safely
extended up to infinity and the result is $\eta_0$-independent.  For
observables quadratic in $k_\perp$, like $y_{23}$, taking into account
the interplay between PT and NP emissions amounts to an effect that is
observable-dependent and goes again as $\sqrt{y_{23}}/Q$.  Finally,
there are observables for which there is a suppression coming from the
vanishing of $\langle g_{1,2}(\phi)\rangle$. This suppression is the
reason why the contribution of the underlying event does not have an
$\eta_0$ enhancement in the case of $T_{m, \cE}$, despite the fact
that $b_{1/2} =0$.
The different ways in which jet hadronization and underlying event
affect the event shape distributions should make it possible to
disentangle the two effects and study them separately.

\section{Conclusions}
We showed that different observables in hadron--hadron collisions have
complementary sensitivities and properties. This makes them a powerful
tool to investigate properties of QCD radiation, in particular the
effects due to jet hadronization and to the underlying event.  These
studies are feasible thanks to the automated resummation tool \caesar.
However, before phenomenological studies can be carried out, resummed
predictions need to be matched with NLO results.  In hadronic
collisions a channel-by-channel matching requires the flavour
information~\cite{Banfi:2006hf} for the fixed-order predictions, which
is unfortunately not currently available to an external user of
hadron--hadron NLO programs~\cite{NLOJET,TriRad}. Work in this
direction is in progress.

\begin{acknowledgements}
  This study was carried out in collaboration with Andrea Banfi and
  Gavin Salam. I thank the organizers of the FRIF workshop for the
  invitation, and for the pleasant and stimulating atmosphere during
  the workshop.
\end{acknowledgements}

\end{document}